\documentstyle[12pt]{article}
\input epsf
\begin{document}
\bibliographystyle{unsrt}
\def\question#1{{{\marginpar{\small \sc #1}}}}
\newcommand{\bra}[1]{\left < \halfthin #1 \right |\halfthin}
\newcommand{\ket}[1]{\left | \halfthin #1 \halfthin \right >}
\newcommand{\be}{\begin{equation}}
\newcommand{\ee}{\end{equation}}
\newcommand{\vsig}{\mbox {\boldmath $\sigma$\unboldmath}}
\newcommand{\vep}{\mbox {\boldmath $\epsilon$\unboldmath}}
\newcommand{\fn}{\frac 1{E^i+M_N}}
\newcommand{\fs}{\frac 1{E^f+M_N}}
\newcommand{\gamQQ}{\Gamma({Q \bar{Q}}_V \rightarrow \gamma + R)}
\newcommand{\brQQ}{b_{rad}({Q \bar{Q}}_V \rightarrow \gamma + R)}
\newcommand{\gamRgg}{\Gamma(R \rightarrow gg )}
\newcommand{\bRgg}{br(R \rightarrow gg )}
\newcommand{\qqbar}{$q \bar{q}~$}
\newcommand{\gsi}{\,\raisebox{-0.13cm}{$\stackrel{\textstyle>}
{\textstyle\sim}$}\,}
\newcommand{\lsi}{\,\raisebox{-0.13cm}{$\stackrel{\textstyle<}
{\textstyle\sim}$}\,} 
\title{\small \rm \begin{flushright} \small{hep-ph/9610426}\\
\small{RAL-96-085}\\
\end{flushright} 
\vspace{0.15cm}
\large \bf   The current picture of Glueballs}
\author{Frank E. Close\thanks{e-mail: fec@v2.rl.ac.uk} \\ 
\small{\it Rutherford Appleton Laboratory}\\
\small{\it Chilton, Didcot, OX11 0QX, England}}
\maketitle
\begin{abstract}
Some recent developments in the phenomenology of the lightest scalar
glueball are summarised. Tools for determining the gluonic content of 
a resonance of known mass, width and $J^{PC}$ from its branching 
fraction in radiative quarkonium decays and production cross section 
in $\gamma \gamma$ collisions are presented.  Two $q\bar{q} - G$ mixing 
schemes for $J^{PC} = 0^{++}$, inspired by the lattice, 
are shown to lead to similar phenomenology 
that may be tested at BEPC and in $\gamma \gamma$ production at LEP2.
\end{abstract}
\vspace{ 2cm}

Rapporteur talk at LEAP96, Dinkelsbuhl, Germany; 27-31 Aug 1996

(based on work with C.Amsler, with G.Farrar and Z.P.Li, and with M.Teper)
\newpage

\section {Introduction}

\hspace*{2em}A quarter of a century after glueballs were first proposed,
it is now looking likely that the scalar glueball has revealed itself, though 
with a novel twist. After searching for a single example, we now have the
luxury of having to choose between two candidates! Furthermore it is of
particular interest to this conference that $p\bar{p}$ annihilation has been
the major player in the experimental developments. 

In the emerging picture
four states are of particular
interest:   
\begin{itemize}
\item 
$f_0(1500)$\cite{lear,cafe95,bugg}
\item
$f_J(1710)$\cite{weing} where $J=0$ or $2$\cite{pdg94}
\item $\xi(2230)$\cite{beijing} 
\item $\eta(1440)$\cite{ishikawa}, now resolved into two pseudoscalars.
\end{itemize}
The interest in these states as glueball candidates is motivated on
both phenomenological and theoretical grounds.  First, phenomenologically,
these states satisfy {\bf qualitative} criteria expected for the production of
glueballs\cite{closerev}:
\begin{enumerate}
\item
Glueballs should be produced in
proton-antiproton annihilation, where the destruction of quarks 
creates opportunity for gluons to be manifested.  This is the Crystal
Barrel \cite{Anis}, and E760 \cite{Hasan1}
production mechanism,
 in which detailed decay systematics of
$f_0(1500)$ have been studied. The empirical situation with regard to
$f_J(1710)$ and $\xi(2230)$ is currently under investigation. The
$\eta(1440)$ is clearly seen in $p\bar{p}$ annihilation\cite{obelixE,cbiota}
\item
Glueballs should be favoured over ordinary mesons in the
central region of high energy scattering processes, away from beam and
target quarks.  The $f_J(1710)$ and possibly the $f_0(1500)$ have been
seen in the central region in $pp$
collisions\cite{Kirk,Gentral}.
\item
Glueballs should be enhanced compared to ordinary mesons in radiative
quarkonium decay.  In fact, all four of these resonances are produced
in radiative $J/\psi$ decay at a level typically of $\sim1$ part per
thousand. 
\end{enumerate}

The latter mechanism has a special role as recent work\cite{cak,cfl96}
 has {\bf quantified} the production rate of conventional mesons
($q\bar{q}$) and glueballs (``$G$") in the radiative decay of vector
quarkonium, as a function of their mass, angular momentum, and width.
If the data on the radiative production of these states are correct, then
\cite{cfl96} finds that     

(i) The $f_0(1500)$ is probably produced at a rate too high to be a
$q\bar{q}$ state.  The average of world data suggests it is a
glueball-$q \bar{q}$ mixture.  

(ii) The $f_J(1710)$ is produced at a rate which is consistent with
it being $q\bar{q}$, only if $J=2$.  If $J=0$, its production
rate is too high for it to be a pure $q\bar{q}$ state but is consistent
with it being a glueball or mixed $q \bar{q}$-glueball having a large
glueball component. 

(iii) The $\xi(2230)$, whose width is $\sim 20$ MeV, is produced at a
rate too high to be a $q\bar{q}$ state for either $J=0$ or $2$.  If
$J=2$, it is consistent with being a glueball.  The assignment $J=0$
would require $Br(J/\psi \rightarrow \gamma \xi) \lsi 3 ~10^{-4}$,
which already may be excluded. 

(iv) The enhancement once called $\eta(1440)$ has been resolved
into two states\cite{cfl96,suchung}.  The higher mass $\eta(1480)$ is dominantly
$s\bar{s}$ with some glue admixture, while the lower state
$\eta(1410)$ has strong affinity for glue.

I shall begin with a brief summary of this quantification. Then I shall review
ideas inspired by the lattice QCD with reference to $f_0(1500)$ and $f_{J=0?}
(1710)$. The latest developments involve mixing schemes motivated by the 
lattice; we shall see that two different schemes\cite{cafe95,wein96}
 have similar
implications which may be tested in experiment.

\section {Production in $\psi \to \gamma R$}

Ref.\cite{cfl96} has used the measured radiative quarkonium production
rates and gamma-gamma decay widths
to make quantitative estimates of the gluonic content of
isosinglet mesons. In particular it applies the relationship of
ref.\cite{cak} between the branching
fraction for a resonance $R$ in radiative quarkonium decay, $\brQQ \equiv
\gamQQ/\Gamma({Q \bar{Q}}_V \rightarrow \gamma + X)$ and its branching
fraction to gluons, $\bRgg \equiv \Gamma(R \rightarrow gg ) /
\Gamma(R\rightarrow \rm{all})$:  
\begin{equation} 
\label{CF}
b_{rad}(Q\bar Q_V\to \gamma +R_J)=
\frac {c_Rx|H_{J}(x)|^2}{8\pi(\pi^2-9)}\frac{m_R}{M_V^2}\Gamma_{tot} br(R_J
\rightarrow gg),
\end{equation}
where $M_V$ and $m_R$ are masses of the initial and final resonances, and
$x \equiv 1-\frac {m_R^2}{M^2_V}$; $c_R$ is a numerical factor and
$H_J(x)$ a loop integral whose magnitude is shown in ref.\cite{cfl96}.
For a
resonance of known mass, total width ($\Gamma_{tot}$), and $J^{PC}$, a
relationship such as eq. (\ref{CF}) would determine $\bRgg$ if 
$\brQQ$ were known.  One may expect
\begin{equation}
\begin{array}{lcl}
br(R[q \bar{q}] \rightarrow gg)& =& 0(\alpha^2_s) \simeq 0.1-0.2\nonumber\\
br(R[G] \rightarrow gg)& \simeq& O(1).\nonumber\\
\end{array}
\end{equation}
Thus knowledge of $\bRgg$ would give quantitative information on the
glueball content of a particular resonance.  Known $q\bar{q}$
resonances (such as $f_2$(1270)) satisfy the former\cite{cfl96}.

In the $x$ regime of immediate interest, $x \sim 0.5 - 0.75$,  one finds
\cite{cfl96} that $\frac{x|H_J|^2}{30-45} \sim O(1)$. This enables us
to manipulate the above into a scaled form that 
exhibits the phenomenological implications immediately.  Specifically, 
for scalar mesons 
\begin{equation} \label{0++}
10^3  br(J/\psi \rightarrow \gamma 0^{++}) = (\frac{m}{1.5\; {\rm GeV}})  
(\frac{\Gamma_{R\rightarrow gg}}{96\; {\rm MeV}})  \frac{x|H_S(x)|^2}{35}.
\end{equation}
This is to be compared with the analogous formula for a tensor meson:
\begin{equation} \label{2++}
10^3  br(J/\psi \rightarrow \gamma 2^{++}) = (\frac{m}{1.5\; {\rm GeV}})  
(\frac{\Gamma_{R\rightarrow gg}}{26\; {\rm MeV}})  \frac{x|H_T(x)|^2}{34}.
\end{equation}
For pseudoscalars we find:
\begin{equation} \label{0-+}
10^3  br(J/\psi \rightarrow \gamma 0^{-+}) = (\frac{m}{1.5\; {\rm GeV}})  
(\frac{\Gamma_{R\rightarrow gg}}{50\; {\rm MeV}})  \frac{x|H_{PS}(x)|^2}{45}.
\end{equation}
Having scaled the expressions this way, because $\frac{x|H_J|^2}{30-45}
\sim O(1)$ in the $x$ range relevant for production of 1.3 - 2.2 GeV
states, we see immediately that the magnitudes 
of the branching  ratios are driven by the denominators 96 and 26 MeV
for $0^{++}$ and $2^{++}$, and $50$ MeV for $0^{-+}$.   Thus if
a state $R_J$ is produced in $J/\psi \rightarrow \gamma X$ at
$O(10^{-3})$ then $\Gamma (R_J \rightarrow gg)$ will typically be of  
the order $100$ MeV for $ 0^{++}$, $O(25 ~ {\rm MeV})$ for $2^{++}$,
and $O(50 ~ {\rm MeV})$ for $0^{-+}$. 

This immediately shows why the $2^{++}$ \qqbar states are prominent: A
$2^{++}$ state with a total width of $O(100 ~\rm{MeV})$ (typical for
$2^{++}$ \qqbar's in this mass range\cite{cafe95,barnes95}) will be
easily visible in $J/\psi \rightarrow \gamma 2^{++}$ with branching
fraction $O(10^{-3})$, while remaining consistent with
\begin{equation} \label{f320} 
br(R[Q\bar{Q}] \rightarrow gg) = 0(\alpha^2_s) \simeq 0.1-0.2.
\end{equation}

Eqs. \ref{0++} - \ref{0-+} not only indicate which $q\bar{q}$ states
will be prominent in $J/\psi \rightarrow \gamma R$, but they also help to
resolve an old paradox concerning $0^{++}$ production.  It was
recognised early on that when the gluons in the absorbtive part of
$J/\psi \rightarrow \gamma gg$ are classified according to their
$J^{PC}$, the partial wave with $2^{++}$ was predicted to dominate.
The waves with $0^{-+}$ and $0^{++}$ were also predicted to be
significant and of comparable strength to one another \cite{bill}.
When extended to include the dispersive part\cite{cak,korner} the
$0^{++}$ was predicted to be prominent over a
considerable part of the kinematic region of interest.  States with $J
\geq 3$ were predicted to have very small rate in this process.
Experimentally, all but one of these appeared to be satisfied.  There
are clear resonant signals in $2^{++}$ and $0^{-+}$, and no
unambiguous signals have been seen with $J \geq 3$.  However no
$0^{++}$ signal had been isolated.  

From our relations above, we see that for a $0^{++}$ to be produced at
the $10^{-3}$ level in $J/\psi$ radiative decay it must either have a
large gluonic content and width $O(100)$ MeV or, if it is a \qqbar
meson, it must have a very large width, $\gsi 500$ MeV.  Taking this
into account, along with the following points, the puzzle of the
absence of $0^{++}$ signal has been resolved:

(i): The width of $^3P_0$ $q\bar{q}$ is predicted to be $\sim 
500$  MeV\cite{cafe95,barnes95}. Thus production at the level
$br(J/\psi \rightarrow \gamma (gg)_{0^+} \sim 10^{-3})$
is consistent with $br(R \rightarrow gg) = 0(\alpha^2_s)
\simeq 0.1-0.2$, but the $\sim 500$ MeV wide signal is smeared
over a large kinematic ($x$) range. 

(ii): The $\sim 100 ~ {\rm MeV}$ wide $f_0(1500)$ signal seen in $J/\psi
\rightarrow \gamma 4\pi$ was originally misidentified as
$0^{-+}$, but is now understood to be $0^{++}$\cite{bugg}.  

(iii): The $f_J(1710)$ which was originally believed to be $J=2$ may
contain a contribution with $J=0$\cite{bugg,pdg94}.

Detailed analysis of the data on $\psi$ radiative decay, when combined
with the above formulae, lead to the conclusions on the gluonic status of
the cited states as listed in the introduction. I refer you to ref.\cite{cfl96}
for more details. Here I would like to report on recent developments arising
from the predictions of lattice QCD concerning the mass and other properties of
the lightest scalar glueball. In particular I shall be interested in the
impact of the $q\bar{q}, J^{PC} = 0^{++}$ nonet being in the vicinity of
the scalar glueball. I shall also end with some speculations on the role that
$S$-wave decays into pseudoscalar meson pairs may have on $0^{++}$ mesons
in the presence of a glueball.

\section{Lattice QCD and Mixing}

Lattice QCD predicts that the lightest
``ideal" (i.e., quenched approximation) glueball be $0^{++}$, with
state-of-the-art mass predictions of $1.55 \pm 0.05$ GeV\cite{ukqcd}
and $1.74 \pm 0.07$ GeV\cite{weing}. A consistent fit to both analyses
gives\cite{ct96} $1.61 \pm 0.07 \pm 0.13$GeV 
where the first error is statistical and the second is systematic.
  That lattice QCD is now
concerned with such fine details represents considerable advance in
the field and raises both opportunity and enigmas. First, it
encourages serious consideration of the further lattice predictions
that the $2^{++}$ glueball lie in the $2.2$ GeV region, and hence
raises interest in the $\xi(2230)$.  Secondly, it suggests that scalar
mesons in the $1.5-1.7$ GeV region merit special attention independent of,
and supplementary to the previous discussion.  The $f_0(1500)$ \cite{cafe95}
and $f_J(1710)$ (if $J=0$)\cite{weinprl} are the two candidates that have
created most interest recently.

A significant new result from the lattice\cite{weinprl} is that the two
body width of a scalar glueball is $\sim O(100)$ MeV and not $\sim
O(1000)$ MeV. In principle the glueball could have been extremely wide
and for practical purposes unobservable. The lattice shows that the
scalar glueball should be a reasonably sharp signal which is an
important guide in helping to eliminate candidates. The width for
decay of the  scalar glueball into pseudoscalar pairs was
predicted\cite{weinprl} to be $108 \pm 28$ MeV. The $f_0(1500)$
has $\Gamma_{tot} = 120 \pm 20$ MeV\cite{pdg96} with the decays into
pseudoscalar pairs comprising $\sim 60$ MeV of this. The $f_J(1710)$
has $\Gamma_{tot} = 140 \pm 12$ MeV. The lattice prediction of the width
guides us towards these
states (if $f_J(1710)$ has $J=0$) but does not of itself discriminate between
them.

Amsler
and Close\cite{cafe95} have pointed out that the $f_0(1500)$ shares
features expected for a glueball that is mixed with the nearby
isoscalar members of the $^3P_0$ $q\bar{q}$ nonet. In particular we noted
that this gives a destructive interference between $s\bar{s}$ and $n\bar{n}$
mixing whereby the $K\bar{K}$ decays are suppressed. This appears to be
the case empirically\cite{landua}. The suppression of $K\bar{K}$ together
with the  {\bf lack} of suppression for $\eta \eta$ is a significant indicator
for a glueball - $q\bar{q}$ mixture\cite{cafe95}.

The properties of the $f_J(1710)$ become central to completing the glueball
picture.  If the $f_J(1710)$
proves to have $J=2$, then it is not a candidate for the ground state
glueball and the $f_0(1500)$ will be essentially unchallenged.
On the other hand, if the $f_J(1710)$ has $J=0$ it becomes a potentially
interesting glueball candidate.  Indeed, Sexton, Vaccarino and
Weingarten\cite{weinprl} argue that $f_{J=0}(1710)$ should be
identified with the ground state glueball, based on its similarity in
mass and decay properties to the state seen in their lattice
simulation. 
The prominent scalar
$f_0(1500)$ was interpreted by ref.\cite{weinprl} 
as the $s\bar{s}$ member of the scalar nonet, however  this identification
does not fit easily with the small $K\bar{K}$ branching ratio and 
the dominant decays to pions.

\subsection{Mixing of Scalar Glueball and Quarkonia}

Whereas the spin of the $f_J(1710)$ remains undetermined, 
it is now clearly established that there are scalar mesons $f_0(1370)$
and $f_0(1500)$ \cite{lear} which couple to $\pi \pi$ and $K\bar{K}$
and so must be allowed for in any analysis of this mass region.

The presence of $f_0(1370), a_0(1450), K_0(1430)$ reinforce the expectation
that a $q\bar{q}$ $^3P_0$ nonet is in the $O(1.3 - 1.7)$GeV mass region.
It is therefore extremely likely that an `ideal' glueball at $\sim 1.6$GeV
\cite{ct96}
will be degenerate with one or other of the $^3P_0$ states given that the widths
of the latter are $O($hundreds MeV). This has
 not been allowed for in any lattice simulation so far.

Ref\cite{cafe95} has considered the phenomenology 
of an ideal glueball lying in 
the midst of the scalar nonet and finds that the ensuing mixings between
the glueball and the nonet lead to a state 
with enfeebled coupling to $K\bar{K}$ (in line with the $f_0(1500)$).
This scenario also leads to significant gluonic 
components in the nearby $n\bar{n}$ and $s\bar{s}$ states. Ref.\cite{cafe95}
proposed that ``if the $f_J(1710)$ is confirmed to have a $J=0$ component in
$K\bar{K}$ but not in $\pi \pi$, this could be a viable candidate for a
$G_0-s\bar{s}$ mixture, completing the scalar meson system built on the
glueball and the quarkonium nonet".

Recently Weingarten\cite{wein96} has proposed
what at first sight appears to be a different mixing scheme 
based on estimates for the mass of the $s\bar{s}$
scalar state in the quenched approximation. Whereas ref\cite{cafe95} supposed 
that the ideal glueball lies within the nonet, ref\cite{wein96} supposed
it to lie above the nonet. I shall now start with the general expressions of
ref\cite{cafe95} and compare the two schemes. This will reveal some rather
general common features.

\subsection{Three-State Mixings}
\label{3mixing}
\hspace*{2em}
An interesting possibility is that three $f_0$'s in the $1.4-1.7$ GeV 
region are admixtures of the three isosinglet states $gg$, $s\bar s$, and
$n\bar n$\cite{cafe95}.  
At leading order in the glueball-$q \bar{q}$ mixing, ref\cite{cafe95}
obtained 
\begin{eqnarray}
\label{mixing}
N_G|G\rangle = |G_0\rangle + \xi ( \sqrt{2} |n\bar{n}\rangle + \omega |s\bar{s}
\rangle) \nonumber \\
N_s|\Psi_s\rangle = |s\bar{s}\rangle - \xi \omega |G_0\rangle \nonumber \\
N_n|\Psi_n\rangle = |n\bar{n}\rangle - \xi  \sqrt{2} |G_0\rangle
\end{eqnarray}
where the $N_i$ are appropriate normalisation factors, $\omega \equiv
\frac{E(G_0) - E(d\bar{d})}{E(G_0) - E(s\bar{s})}$ and the mixing parameter
$\xi \equiv \frac{\langle d\bar{d}|V|G_0\rangle}{E(G_0) - E(d\bar{d})}$. The
analysis of ref\cite{cfl96}
suggests that the $gg \to q\bar{q}$ mixing amplitude manifested
in $\psi \to \gamma R(q\bar{q})$ is $O(\alpha_s)$, so that qualitatively
$\xi \sim O(\alpha_s) \sim 0.5$. Such a magnitude implies significant mixing
in eq.(\ref{mixing}) and is better generalised to a $3 \times 3$ mixing matrix.
Ref.\cite{wein96} defines this to be
$$
\begin{array}{c c c c}
m_G^0 & z & \sqrt{2} z \\
z & m_s^0 & 0  \\
\sqrt{2} z & 0 & m_n^0 \\
\end{array}
$$ 
where
$z \equiv \xi \times (E(G_0) - E(d\bar{d}))$ in the notation of 
ref.\cite{cafe95}.

Mixing based on lattice glueball masses lead to two classes of solution
of immediate interest:  

\noindent (i)$\omega \leq 0$, corresponding to $G_0$ in the midst
of the nonet\cite{cafe95} 

\noindent (ii)$\omega > 1$, corresponding to $G_0$ above the
$q\bar{q}$ members of the nonet\cite{wein96}. 

The model of Genovese\cite{genovese} is a particular case where
$\xi \to 0; \omega \to \infty$ with $\xi \omega \to 1$.

We shall denote the three mass eigenstates by $R_i$ with $R_1=f_0(1370)$,
$R_2=f_0(1500)$ and $R_3=f_0(1710)$, and the three isosinglet states 
$\phi_i$ with $\phi_1=n\bar n$, $\phi_2=s\bar s$ and $\phi_3=gg$ so
that $R_i=f_{ij}\phi_i$.  

There are indications from lattice QCD that the scalar $s\bar{s}$ state,
in the quenched approximation, may lie lower than the scalar glueball
\cite{lacock96,wein96}. Weingarten\cite{wein96} has constructed a
mixing model based on this scenario. The input ``bare" masses are
$m_n^0 = 1450; m_s^0 = 1516; m_G^0 = 1642$ and the mixing strength
$z \equiv \xi \times (E(G_0) - E(d\bar{d})) = 72$ MeV. The
resulting mixtures are
$$
\begin{array}{c c c c}
&f_{i1}^{(n)} & f_{i2}^{(s)} & f_{i3}^{(G)}\\
f_0(1370) & 0.87 &  0.25 & -0.43\\
f_0(1500) & -0.36 &  0.91 & -0.22\\
f_0(1710) & 0.34 & 0.33 & 0.88\\
\end{array}
$$ 

It is suggested, but not demonstrated, that the decays of the
$f_0(1500)$ involve significant destructive interference between its
gluonic and $s\bar{s}$ components whereby the $K\bar{K}$ suppression and
$2\pi$, $4\pi$ enhancements are explained. The disparity between $K\bar{K}$
suppression and the {\bf strong} coupling to $\eta \eta$ remains an open
question here.

Recent data on the decay $f_0(1500) \to
K\bar{K}$\cite{landua} may be interpreted within the scheme of
ref\cite{cafe95} as being consistent with the $G_0$ lying between
$n\bar{n}$ and $s\bar{s}$ such that the parameter $\omega \sim
-2$. (In this case the $\eta \eta$ 
production is driven by the gluonic component of
the wavefunction almost entirely,see ref\cite{cafe95}).
If for illustration we adopt $\xi
=0.5 \sim \alpha_s$, the resulting mixing amplitudes are 
$$
\begin{array}{c c c c}
&f_{i1}^{(n)} & f_{i2}^{(s)} & f_{i3}^{(G)} \\
f_0(1370) & 0.86 &  0.13 & -0.50\\
f_0(1500) & 0.43 & - 0.61 & 0.61\\
f_0(1710) & 0.22 & 0.76 & 0.60\\
\end{array}
$$

The solutions for the lowest mass state in the two schemes
are similar, as are the relative
phases and qualitative importance of the $G$ component in the high
mass state.  Both solutions exhibit destructive interference between
the $n\bar{n}$ and  $s\bar{s}$ flavours for the middle state. 

This
parallelism is not a coincidence.
A general feature of this three way mixing is that in the limit of
strong mixing the central state tends towards flavour octet with the
outer (heaviest and lightest) states being orthogonal mixtures of
glueball and flavour singlet, namely 

$$
\begin{array}{c c}
f_0(1370) & |q\bar{q}(\bf{1})\rangle - |G\rangle \\
f_0(1500) & |q\bar{q}(\bf{8})\rangle + \epsilon |G\rangle \\
f_0(1710) & |q\bar{q}(\bf{1})\rangle + |G\rangle \\
\end{array}
$$
where $\epsilon \sim \xi^{-1} \to 0$.

In short, the glueball has leaked away maximally
to the outer states even in the case (ref\cite{cafe95}) where the bare glueball
(zero mixing) was in the middle of the nonet to start with. The leakage into
the outer states becomes significant once the mixing strength (off diagonal
 term in the mass matrix) becomes comparable to the mass gap between glueball
and $q\bar{q}$ states (i.e. either $\xi \geq 1$ or $\xi \omega \geq 1$).
 Even in the zero width approximation of ref\cite{cafe95}
this tends to be the case and when one allows for the widths being of 
$O(100)$MeV while the nonet masses and glueball mass are spread over only
a few hundred MeV, it is apparent that there will be considerable leakage
from the glueball into the $q\bar{q}$ nonet. It is for this reason,
 {\it inter alia}, that the output of refs\cite{cafe95} and \cite{wein96}
are rather similar. While this similarity
may make it hard to distinguish between
them, it does enable data to eliminate the general idea should their
common implications fail empirically.

If we make the simplifying assumption that the photons couple to the
$n\bar{n}$ and  $s\bar{s}$ in direct proportion to the respective
$e_i^2$ (i.e. we ignore mass effects and any differences between the
$n\bar{n}$ and $s\bar{s}$ wavefunctions), then the corresponding two
photon widths can be written in terms of these mixing coefficients:
\begin{equation}\label{mixings}
\Gamma(R_i)=|f_{i1}\frac {5}{9\sqrt{2}}+f_{i2}\frac {1}{9}|^2 \Gamma,
\end{equation}
where $\Gamma$ is the $\gamma\gamma$ width for a $q\bar q$ system
with $e_q=1$.  One can use eq. (\ref{mixings}) to evaluate the
relative strength of the two photon widths for the three $f_0$ 
states with the input of the mixing coefficients\cite{cfl96}.
  These are (ignoring
mass dependent effects) 
\begin{equation} \label{acgamma}
f_0(1370) : f_0(1500) : f_0(1710) \sim 12:1:3
\end{equation}
in the Amsler - Close scheme \cite{cafe95} to be compared with
\begin{equation} \label{weingamma}
f_0(1370) : f_0(1500) : f_0(1710) \sim 13:0.2:3
\end{equation}
in Weingarten\cite{wein96}. At present the only measured $\gamma \gamma$ width
in this list is that of the $f_0(1370) = 5.4 \pm 2.3$ keV\cite{pdg94}.
Using this to normalise the above, we anticipate $f_0(1500) \to \gamma
\gamma \sim 0.5$ keV \cite{cafe95} or $\sim 0.1$ keV \cite{wein96}.
Both schemes imply $\Gamma(f_0(1710) \to \gamma \gamma) =
1-2$ keV. 

This relative ordering of $\gamma \gamma$ widths is a common feature
of mixings for all initial configurations for which the bare glueball
does not lie nearly degenerate to the $n\bar{n}$ state.  As such, it
is a robust test of the general idea of $n\bar{n}$ and $s\bar{s}$
mixing with a lattice motivated glueball.  If, say, the $\gamma
\gamma$ width of the $f_0(1710)$ were to be smaller than the
$f_0(1500)$, or comparable to or greater than the $f_0(1370)$, then
the general hypothesis of significant three state mixing with a
lattice glueball would be disproven. The corollary is that qualitative
agreement may be used to begin isolating in detail the mixing pattern. 

The production of these states in $\psi \to \gamma f_0$ also shares some common
features in that $f_0(1710)$ production is predicted to dominate. The
analysis of ref.\cite{cfl96} predicts that
\begin{equation}
br(J/\psi \to \gamma \Sigma f_0) \geq (1.5 \pm 0.6) \times 10^{-3}.
\end{equation}
In \cite{cafe95} the $q\bar{q}$ admixture in the $f_0(1500)$ is nearly
pure flavour octet and hence decouples from $gg$. This leaves the
strength of  $br(J/\psi \to \gamma f_0(1500))$ driven entirely by its
$gg$ component at about $40\%$ of the
pure glueball strength. This appears to be consistent with the mean of the 
world data (for details see ref.\cite{cfl96}).

Thus, in conclusion, both these mixing schemes imply a similar hierachy
of strengths in $\gamma \gamma$ production which may be used as a 
test of the general idea of three state mixing between glueball and
a nearby nonet. Prominent production of $J/\psi \to \gamma f_0(1710)$
is also a common feature.  When the experimental situation clarifies
on the $J/\psi \to \gamma f_0$ branching fractions, we can use the
relative strengths to distinguish between the case where the glueball
lies within a nonet, ref\cite{cafe95}, or above the $s\bar{s}$ member,
ref\cite{wein96}. 

In the former case this $G_0 - q\bar{q}$ mixing
gives a destructive interference between $s\bar{s}$ and $n\bar{n}$
whereby decays into $K\bar{K}$ are suppressed.
However, even in the case where the $s\bar{s}$ lies below the $G_0$
we expect that there will be
$K\bar{K}$ destructive effects due to mixing with not only
with the $s\bar{s}$ that lies
below $G_0$ (as in ref.\cite{wein96}) but also 
with a radially excited $n\bar{n}$
lying above it (not considered in ref.\cite{wein96}). Unless $G_0$ mixing with
the radial state is much suppressed, this will give a similar pattern to that
of ref.\cite{cafe95} though with more model dependence due to the differing
spatial wavefunctions for the two nonets. In the final section I shall
assume that the $G_0 - q\bar{q}$ mixing leads to $n\bar{n} - s\bar{s}$
in the wavefunction.

\section{Glue, $q\bar{q}$ and Mesons: A Hierarchy}

Scalar quarkoonium 
mesons are P-wave $q\bar{q}$ but decay in S-wave to pseudoscalar 
meson pairs. Tornqvist\cite{torn95}
 has argued that this can distort the meson nonet 
considerably, in particular dragging the $s\bar{s}$ and $I=1$ bare states
down to the vicinity of $K\bar{K}$ threshold (to be identified with
the $f_0(980)$ and $a_0(980)$ respectively). His analysis has not included
the possible role of a primitive glueball.

On the other hand the  analyses in refs.\cite{cafe95,wein96}
have considered mixing of $gg$
and $q\bar{q}$ without inclusion of meson pairs. It is amusing to consider
the qualitative picture that might emerge when the parton
\cite{cafe95} and meson\cite{torn95} are combined.

The main question is how the glueball couples to the hadronic sector.
In ref.\cite{cafe95} we argued that mixing into (nearby) $q\bar{q}$
may play a leading role, at least for scalar glueball. This will
cause a suppression of $K\bar{K}$ if the mixing is dominantly with nearest 
neighbours: this is clear in the scheme of ref\cite{cafe95} where the 
glueball lies in the middle of a nonet, but will also happen if the glueball 
lies above the $s\bar{s}$ as proposed by Weingarten\cite{wein96}. The point is
that in the $q\bar{q}$ spectroscopy (where $N$ denotes
the radial excitation quantum number) the $N(s\bar{s})$ will lie
below, and near to, $(N+1)(n\bar{n}$). Consequently we may expect
that in general a glueball will lie
either between $n\bar{n}$ and $s\bar{s}$ of a single multiplet, or
between $s\bar{s}$ and $n\bar{n}$ of adjacent multiplets. While the
overlapping of multiplets for the higher mass $2^{++}$ (which can also involve
$^3F_2$ as well as radial excitation of $^3P_2$) may break this flavour 
ordering, it is a rather general feature in potential models for the
1.5 GeV region for scalars.

If we adopt this $K\bar{K}$ suppression, we find that (at least for
the $q\bar{q}$ decays) the $\eta \eta$ and $\eta' \eta'$ are also
suppressed. The $\pi \pi$ threshold is far away from the 1.6 GeV region
and the intrinsic
Clebsch Gordan coefficient coupling $\frac{1}{\sqrt{2}} (n\bar{n} - s\bar{s})
cos \theta + sin \theta G_0$ to $\pi \pi$ 
is also somewhat reduced ($cos\theta / 2$) due to this mixing.
A major effect may be expected from the $\eta \eta'$ threshold. This channel
couples strongly to the $n\bar{n} - s\bar{s}$ mixture\cite{cafe95} and
possibly also to $G_0$ through glue content in the $\eta$ system.
Furthermore the $\eta \eta'$ 
threshold of $1510$ MeV is near to the glueball mass, 
according to lattice QCD\cite{ukqcd,ct96,weinprl}.
Thus it may be natural to find a strong attraction of the glueball to the 
1500 MeV mass region. It may be interesting to extend the $G_0 \to q\bar{q}$
mixing analysis of ref\cite{cafe95} to include the $q\bar{q} \to 0^- 0^-$
unitarisation of ref.\cite{torn95}. This leads to a more 
complete description of the scalar mesons and S-wave thresholds in the 1
to 2 GeV region\cite{pen96}. The implications for $\gamma \gamma$ couplings, in
particular, could then enable the role and parameters
of the scalar glueball to be quantified.

In parallel a study of flavour decays based on the lattice results of
ref\cite{weinprl} and the mixing schemes of section 3.2 is warranted.
The focus now is on ways to disentangle the glueball dynamics from the
scalar mesons in the 1.4 to 1.8 GeV region through a dedicated programme
concentrating on the $f_0(1500)$ and $f_J(1710)$ in particular. The emergent
data are remarkably consistent with models based on lattice QCD. This
is real progress compared even to two years ago and is due in no small
part to the remarkable data that have emerged from $p\bar{p}$ annihilation
at LEAR.

\section{Acknowledgements}
This review owes much to my collaborations with C.Amsler, G.Farrar, Z.P.Li
and M.Teper and it is those original works that should be cited. 
I am also indebted to colleagues at LEAP96 for their comments
during and after the conference. This work is partially
supported by the European Community Human Mobility Program Eurodafne,
Contract CHRX-CT92-0026.

\end{document}